\documentclass[12pt]{article}
\usepackage{amssymb,amsmath,esint}
\begin{document}


\begin{titlepage}

                             \begin{center}

\Large\bf Moduli of curve families and (quasi-)conformality of power law entropies.

                            \vspace*{2.5cm}

              \normalsize\sf    NIKOS \ KALOGEROPOULOS $^\dagger$\\

                            \vspace{0.2cm}
                            
 \normalsize\sf Weill Cornell Medicine \  - \ Qatar\\
 Education City,  P.O.  Box 24144\\
 Doha, Qatar\\

                            \end{center}

                            \vspace{2.5cm}

                     \centerline{\normalsize\bf Abstract}
                     
                           \vspace{3mm}
                     
\normalsize\rm\setlength{\baselineskip}{18pt} 

We present aspects of the moduli of curve families on a metric measure space which may prove useful in calculating, or in providing bounds to, non-additive entropies having a power-
law functional form. We use as paradigmatic cases the calculations of the moduli of curve families for a cylinder and for an annulus in $\mathbb{R}^n$. 
The underlying motivation for these studies is 
that the definitions and some properties of the modulus of a curve family resembles those of the Tsallis entropy, when the latter is seen from a micro-canonical viewpoint. We comment on 
the origin of the conjectured invariance of the Tsallis entropy under M\"{o}bius transformations of the non-extensive (entropic) parameter. Needing techniques applicable to both locallly 
Euclidean and fractal classes of spaces, we examine the behavior of the Tsallis functional, via the modulus, under quasi-conformal maps. We comment on properties of such maps and 
their possible significance for the dynamical foundations of power-law entropies.\\

                           \vfill

                                              \hspace*{21mm} Quasi-conformal maps, Quasi-M\"{o}bius maps.   \\
                             
                             \vfill

\noindent\rule{8cm}{0.2mm}\\  
   \noindent $^\dagger$ \small\rm E-mail: \ \  nik2011@qatar-med.cornell.edu\\

\end{titlepage}
 

                                                                                \newpage                 

\rm\normalsize
\setlength{\baselineskip}{18pt}

\section{ Introduction.}

The Havrda-Charv\'{a}t \cite{HC}/Vajda \cite{Vaj}/Dar\'{o}czy \cite{Dar}/Lindhard-Nielsen \cite{LN}/Cressie-Read \cite{CR, RC}/Tsallis \cite{T1, T-book}  
entropic functional, henceforth to be called just ``Tsallis entropy" for brevity, is a one-parameter family of entropic functionals $\mathcal{S}_q$ parametrized 
by the non-extensive (entropic) parameter $q\in\mathbb{R}$, which is given (for a discrete set of outcomes parametrized by elements $i$ of an index set $I$ 
with corresponding probabilities $p_i$) by 
\begin{equation}
       \mathcal{S}_q [ \{ p_i \} ] \ =  \  \ -  \ k_B \cdot \frac{1}{q-1}  \left( 1 - \sum_{i\in I} \ p_i^q  \right)
\end{equation}
where $k_B$ stands for Boltzmann's constant.                                                      
There has been a recent proposal \cite{WW} about extending the domain of $q$ to $q\in\mathbb{C}$. Such a suggestion, for the purposes of this work, is notable and 
suggestive, even if not necessarily of great use for the calculations  or arguments explicitly needed here.\\

 The extension of (1) to continuous sample spaces $\Omega$, which are usually taken to be 
Riemannian manifolds ($\mathbf{M, g}$) with occasionally additional properties, such as a symplectic structure \cite{Arnold, Zehn}, seems not to be entirely trivial. 
The subject is still unsettled, in our opinion, in light of a recent controversy \cite{Abe1, Andre, Abe2, BOT, BL, LB, PR}. 
This controversy is related to the fact that even the Boltzmann/Gibbs/Shannon entropy $\mathcal{S}_{BGS}$ which for continuous systems $\Omega$ 
with corresponding measure $d\mu_\Omega$ is given by 
\begin{equation}
    \mathcal{S}_{BGS} [\{p_i \}] \ = \  - \ k_B \ \int_{\Omega} \  \rho \log\rho \ d\mu_{\Omega}    \nonumber
\end{equation}
 has, implicitly, two features \cite{Balian, Lesne}  which may become problematic in generalized functionals such as (1): 
the first is coordinate dependence of $\mathcal{S}_{BGS}$. 
Obviously, the parametrization of the system should not matter when one analyzes physical aspects of such a system. 
To address this, one actually calculates not the ``absolute" entropy but a  relative entropy,
akin to the Kullback-Leibler divergence. This, in turn, brings in a second problematic feature: such a relative entropy needs a background distribution with respect to 
which it has to be calculated. How does one choose such a background distribution and its physical significance can be non-trivial. 
Such issues have been largely dealt with satisfactorily in the case of  $\mathcal{S}_{BGS}$. However they are still present and largely unaddressed for 
the case of the continuum analogue of (1). Due to the fact that such issues may significantly affect the mathematical consistency and physical basis of the 
thermodynamic  predictions based on the continuum analogue of (1), they need to be clarified. 
As an example of the technical problems posed by extending naively (1) to the continuum case, one can point out that in such an expression, if one chooses a background 
measure as in the case of $\mathcal{S}_{BGS}$, such a background measure  remains explicitly present in the thermodynamic limit, 
due to the power law (as opposed to the logarithmic, for $\mathcal{S}_{BGS}$) functional dependence of (1). This is a point that the reader of any work relying on 
the subsequent expression (2) below should have in mind, as it potentially pertains to the applicability/validity of (2) on which the contents of the present work  
rely. \\    
      
Having stated the above, and to be able to proceed without confronting 
such an issue whose resolution may be important but will lead us too far in a different direction from our intended goals, 
we will assume for the purposes of this  work that the naive extension of (1) to continuous cases, which has been 
used extensively in the past in determining numerous properties of the Tsallis entropy,  is valid and is given by
\begin{equation}         
       \mathcal{S}_q [\rho ] \ = \  - \ k_B \cdot \frac{1}{q-1} \left( 1 - \int_{\mathbf{\Omega}} \ [\rho (x) ]^q \ dvol_{\Omega} \right)
\end{equation}
where $\rho$ stands for the Radon-Nikodym derivative of the probabilty density function of the effective process with respect to the volume element \ 
$dvol_{\Omega}$ \  of \ $\Omega$. \\

It is well-known that the explicit analytic calculation of $\mathcal{S}_q$ for specific systems is a difficult task. In particular, the determination of $\rho$ in (2) from dynamical 
principles is, arguably, the most important step in connecting the microscopic to the macroscopic properties of a system of many degrees of freedom.        
Such a step is accomplished in the case of the Boltzmann/Gibbs/Shannon entropy $\mathcal{S}_{BGS}$ by invoking additional ``reasonable" assumptions such as the 
molecular chaos hypothesis or the mixing behavior of the pertinent measure densities $\rho$ in phase space. Given that  non-additive entropies, 
such as $\mathcal{S}_q$ in particular, claim to describe non-ergodic systems, it becomes far from obvious how to determine such a $\rho$ from the dynamics of 
the underlying microscopic system and then how to actually calculate (2) for such a $\rho$ in concrete models. One motivation for the present work is to see on 
whether there are mathematical constructions that essentially ``derive" such a $\rho$ from the phase space portrait of the underlying Hamiltonian system.   
Another motivation is to  look into a conjectured invariance related to  different values of the non-extensive parameter $q$. Such an invariance seems to 
have been observed in data fittings which are conjecturally described by the equilibrium distributions (the $q$-exponentials) resulting from applying the maximum entropy 
principle to $\mathcal{S}_q$. It appears that there are systems for which (2) remains invariant under the transformations 
\begin{equation}
       q \  \ \mapsto \  \ 2-q     \hspace{25mm}  q \   \  \mapsto \  \  \frac{1}{q} 
\end{equation}
The transformations (3) are a set of generators of M\"{o}bius transformations on the complex plane, assuming $q\in \mathbb{C}$. More generally, 
it may be worth examining the behavior of (2) under properties resembling but also generalizing such M\"{o}bius
transformations. Such behavior may also be shared by  other entropic functionals of power-law form that have been studied recently.\\  

In this line of investigation, we encountered  a fundamental construction in geometric function theory called the moduli of families of curves. For our purposes, we imagine such 
curves to be trajectories of the microscopic Hamiltonian system whose thermodynamic behavior is conjecturally captured by $\mathcal{S}_q$.  The actual roots of the moduli of 
curves construction can be traced back to the development of electromagnetic theory. Some details of this connection and some references are provided in Subsection 3.5.
The construction and properties of the moduli of curve families  have several commonalities with the calculation of (2) as we explicitly point out in Subsection 2.3. 
The present work deals with  aspects of the moduli of curves which may be possibly pertinent, in our opinion, to the dynamical foundations of non-additive entropies. 
We are also looking into general features of maps leaving (almost) invariant the moduli of curves families and their connection with established physical and formal aspects of (2).  \\

In Section 2, we present a few details about the moduli of curve families and also state two of the very few explicit examples where the computation of such moduli  is feasible, 
from the viewpoint of their relevance to $\mathcal{S}_q$. In Section 3, and in the same spirit, we discuss a few concepts of the ``quasi-world" and ``coarse" geometry, 
emphasizing quasi-conformal maps, and discuss the invariance  of the moduli of curve families  under such homeomorphisms. 
In Section 4 we present some conclusions and an outlook toward future developments in this direction.\\

                                                       \vspace{0mm}


\section{ Moduli of curve families.}

                                                       \vspace{1mm}

\subsection{Motivation and definition.} 
There are several ways that one can motivate the definition of the moduli of curve families, each representing a 
particular viewpoint. For our entropic-related purposes, consider, for example,  an autonomous  Hamiltonian system of many degrees of 
freedom whose thermodynamic behavior we want to describe. If the system is ergodic, something quite hard to actually verify in concrete cases, 
then the employed measure is, up to a normalization constant,  the infinitesimal area measure $dvol_E$ of the constant energy $E$ 
hypersurface $\mathcal{M}_E$ of the phase space $\mathbf{M}$, in a micro-canonical treatment. The entropy $\mathcal{S}_q$ conjecturally describes 
non-ergodic systems, among many other things \cite{T-book}, so one can generally consider as an effective measure a deformation of $dvol_E$  expressed by some function 
$\rho :\mathbf{M} \rightarrow \mathbb{R}_+$ as
\begin{equation}   
    d\mu_E \ = \ \rho \ dvol_E
\end{equation}
This expression tacitly assumes that the departure from ergodicity can be expressed through a measure $d\mu_E$ which is absolutely continuous 
with respect to the hypersurface volume $dvol_E$. Obviously $\rho$ is the normalization constant in the micro-canonical analysis of the system.  
We discussed in \cite{NK1, NK2, NK3, NK4}, directly or indirectly, some geometric consequences of such a deformation $\rho$ pertinent to $\mathcal{S}_q$.\\
 
The question that naturally arises is how to actually determine such a $\rho$. Even though the  answer is clearly model-dependent, we may wish to find 
some common features that are shared by such answers for a variety of systems. The most dynamic scenario, following the spirit of L. Boltzmann and A. Einstein, 
would be to derive such a $\rho$  from the dynamics of the underlying, ``microscopic", Hamiltonian system. 
 On the opposite end of the spectrum, a ``least dynamical" approach relying on statistical inference, rather than dynamics, is expressed   
by the ``maximum entropy principle", as has been advocated by several people, most notably by E.T. Jaynes \cite{Jaynes}. 
We follow, as much as we can in this work,  the dynamical approach. Dynamics is considered by some people (like L.D. Landau, A.I. Khintchin etc) 
as largely irrelevant to Statistical Mechanics whose validity, as they argue, relies on the large number of degrees of freedom of the underlying system.
The issue remains largely unsettled even to this day \cite{Vulp}.  Our viewpoint is that  when we depart from the well-established realm of phenomena 
described through $\mathcal{S}_{BGS} $, features of the underlying dynamics may be helpful understanding which, if any, of the proposed entropic 
functionals can be used for the study of the collective properties of the system at hand. \\       

To advance the viewpoint relying on the underlying dynamics, some connection between the underlying dynamics and the distribution $\rho$ is needed. 
A well-known way to encode this goes as follows: consider a collection  of $N\in\mathbb{N}$ points in a subset of phase space $\Omega \subset \mathbf{M}$ 
denoted by $\mathbf{X}^{(i)}, \  i=1,2,\ldots N$. The superscript is labelling the different points, rather than being an explicit coordinate. 
A probability density on $\Omega$ expressing the collective dynamics of  $\mathbf{X}^{(i)}$ can be given as the mean of the contributions of all such points:
\begin{equation} 
   \rho (\Omega ) \  = \lim_{N\rightarrow \infty}  \frac{1}{N} \  \sum_{i=1}^N \  \delta (\Omega - \mathbf{X}^{(i)} )
\end{equation}
where $\delta$ stands for the Dirac delta function. 
More specifically, if these points represent the initial conditions, indicated by the subscript $0$ of a dynamical system whose evolution (trajectories)  
is given by the  flow $\varphi_t, \ t\in\mathbb{R}_+$, then at at later time
\begin{equation}  
       \rho_t (\Omega) \ = \ \lim_{N\rightarrow\infty} \frac{1}{N} \  \sum_{i=1}^N \ \delta (\Omega - \varphi_t (\mathbf{X}_0^{(i)})) 
\end{equation}
An obvious question that arises is whether (6) is meaningful, and if so, to what extent and under what conditions.  A more practical aspect is how someone can actually compute 
an expression like (6) for anything beyond simple ``toy" models of very few degrees of freedom. None of these questions seems to have obvious or simple answers for 
even ``semi-realistic" physical systems.\\  

As seen in Hamilton's equations, the evolution of a system under given initial conditions is a curve $\gamma$ in phase space $\mathbf{M}$.
This curve should not be self-intersecting, can ``turn", ``twist" and more generally have a complicated geometric behavior in $\mathbf{M}$. 
However all such curves have a length $L(\gamma)$ which is defined with respect  to the quadratic, in canonical momenta, part of the underlying Hamiltonian $\mathcal{H}$ , 
which can be used  as an effective Riemannian metric $\mathbf{g}$. It should be noticed that in Hamiltonian Mechanics, there is no a priori ``natural" metric on phase 
space $\mathbf{M}$ \cite{CPC, Klauder}.   One can, of course, choose many metrics in a geometric treatment of Mechanics, but none of them is superior to the other choices generically. 
The phase space is just a symplectic manifold  with or without additional structure, depending on the particular model which is analyzed \cite{Arnold, Zehn}.
The length $L (\gamma)$ of a phase space trajectory is assumed to be finite, locally at least or, in technical terms, $\gamma$ to be locally rectifiable \cite{Fed, AGS}. 
Note that this assumption excludes from phase portraits showflakes, or similar, curves  which are not locally rectifiable, but this should not be 
a substantial restriction due to the sufficiently regular behavior of the Hamiltonian evolution that excludes such curves being trajectories of physical systems.
It can however pose problems in the thermodynamic limit as there is no guarantee about the regularity of the curves in this limit, without imposing additional conditions.  
The profile of the images all such curves $\gamma$ in $\mathbf{M}$ should give rise to $\rho$ as (6) formally indicates.  \\

Another way to construct such $\rho$ is to demand that such $\rho$ should not decrease the length of the evolution trajectories in phase space. 
Such $\rho$ is a density which should be playing an important role in non-ergodic systems, such as the ones that are conjecturally described by $\mathcal{S}_q$.  
The ``density" $\rho$ can be arbitrarily close or even exactly zero at subsets of phase space. 
In such cases however, it will not really contribute anything to the statistical description of the system. If we assume that the phase space $\mathbf{M}$ is compact, we can exclude from 
our considerations subsets at which $\rho \neq 0$, but $\rho$ is small. This is akin to the thick-thin decomposition extensively used in hyperbolic geometry \cite{Petronio}.
 We arbitrarily choose a non-zero lower bound for $\rho$ that allows us to exclude regions of phase space that 
are ``rarely" visited by the system. Then we can re-normalize $\rho$ with respect to this lower bound. To formalize this idea,  assume that $\gamma (s)$ is arc-length parametrized by 
$s\in [0, L(\gamma)]$ and require that 
\begin{equation}  
    \int_0 ^{L(\gamma)} \rho (\gamma(s)) \ ds \ \geq \ 1  
\end{equation}
Formally $\rho$ is assumed to be a Borel function $\rho: \mathbf{M} \rightarrow [0, \infty]$. Functions $\rho$ obeying  (7) are called admissible functions (or metrics)
for the curve $\gamma$ or, more precisely,  for the set of curves $\mathcal{A}(\gamma)$.  
Finding the maximum of the Tsallis entropy (2) over all such $\rho$ obeying (7) amounts to determining
\begin{equation}
           \inf_\rho \int_{\mathbf{M}} [\rho(x)]^q \  dvol_{\mathbf{M}}  
\end{equation}
in this language. In no statements above  have we used that ($\mathbf{M, g}$) should be a Riemannian manifold. Hence (7), (8) can be formulated for a general metric measure space 
($\mathfrak{X}, d, \mu$) without any assumptions about its regularity or other properties beyond the ones needed to define the existence and length of a 
rectifiable curve and a Borel measure $dvol_M$ or, more generally, $\mu$ \cite{Fed, AGS}. 
Then one formally defines, for $q \geq 1$ and all $\rho$ obeying (7), the $q$-modulus of the family of curves 
$\mathcal{A} (\gamma )$  by
\begin{equation}
   \mod_{\!\! q}  \mathcal{A} \ = \  \inf_\rho \int_\mathfrak{X} \  [\rho (x)]^q \  d\mu 
\end{equation}
We point out, for completeness, that, by definition, the modulus of unrectifiable curves is zero and that the modulus of a family involving a constant curve is infinite. 
We see from the (9), that the maximization of $\mathcal{S}_q$ in a micro-canonical setting, is akin to the computation of the $q$-modulus of a curve family.  
For the computation to be  meaningful in the case of Hamiltonian systems, the curves $\gamma \in \mathcal{A}$ should be the evolution trajectories  of the Hamiltonian 
system in $\mathfrak{X}$  under various initial conditions. We should also re-iterate the restriction that $q\geq 1$ applicable in (9), but not in (2). \\

We would like to single out \cite{Vai, Vuor, HeinKosk, Hein, MRSY, HKST, Kosk} among the numerous excellent references where many aspects, 
mostly from a metric viewpoint and applicable in any dimension, of the moduli of curve families are presented very clearly and in far greater detail than our 
minimalistic  presentation in the sequel. \\  


\subsection{Further interpretation of the modulus.}

Consider a non-ergodic system on $\mathfrak{X}$. If  many curves, or their segments, belonging to the evolution of the dynamical system pass through a neighborhood  
$U_x$ of $x\in \mathfrak{X}$  then the ``density" function $\rho$ will have a large value in $U_x$. Otherwise its value will be smaller, but certainly $\rho \geq 0$. If someone 
 uses as weight the function $\rho$, and calculates the weighted length of any curve in $U_x$, then the result will be (7) but with the upper limit of integration modified 
to be the length of the curve in $U_x$, which is certainly smaller than $L(\gamma)$.  If the system were ergodic, then $\rho$ would be constant and (7) would trivially hold 
as equality. Hence a condition like (7) is necessary to provide a lower bound to how small $\rho$ can be, in order to avoid allowing for (9) to be saturated by the trivial distribution 
$\rho = 0$. The fact is that even though (7) is a simple relation, it is  related to the underlying dynamics, and this makes (7) very desirable.\\

A second point is that there is no requirement in (7) or (9) for $\rho$ to be continuous. It actually turns out that  using lower semi-continuous, rather than continuous, 
as admissible functions is sufficient,  without sacrificing any generality in the scope of the arguments. 
This is initially assumed as a technical point, in order to be general in a way that would allow 
 theorems pertaining to the modulus but having a wide scope to be proved. Its significance for Physics however is the following: 
 it is indeed true that $\rho$ in phase space should be continuous. 
 Discontinuities  may arise for at  least two reasons: one is the thermodynamic limit. It is easy to see that, in general, limits of sequences of continuous functions are not 
 continuous, unless we impose additional conditions \cite{Rudin} on the elements of such sequences or on the definition of the limit. Another reason is that it may not
 in general be true that $\rho$ should be everywhere regular with respect to the underlying measure/volume $d\mu$. The measure \ $\rho \ d\mu$ \ may become singular in parts of the 
 phase space, as in the case of the Sinai-Ruelle-Bowen (SRB) measures' projection on the stable manifolds of a hyperbolic system. The SRB measures may apply to 
 dissipative, hence non-Hamiltonian systems, however it may be desirable to keep a maximal level of flexibility in accommodating different classes of dynamical systems 
 in the definitions that we use. In such cases the density $\rho$ may become distributional or not exist at all, depending on the circumstances. 
 A discontinuity in $\rho$ may help determine such a potentially unexpected, but not disastrous, in the present framework, behavior.\\
 
A more geometric meaning of (9), which is still closely associated to the Tsallis entropy (2),  goes as follows: Assume that we analyze a system by using a particular metric 
in its phase space. Such a choice of  metric may not be canonical \cite{Klauder}, but may be useful. 
We want to know how predictions would change if such a metric is modified/perturbed. This is spirit quite common in Quantum Physics, for instance.
We start from a classical system which is assumed to be reasonably well-understood. Then attempt to we quantize it using our favorite method. In the process it may turn out that, 
for whatever reasons, some of the initial assumptions, however reasonable and successful may have been for the classical system, may have to be modified for the quantum case.
Just consider the Coleman-Weinberg mechanism and dimensional transmutation \cite{IZ} which is the prototypical example of dynamical symmetry breaking in field theories. One usually 
starts from a potential having a simple monomial/polynomial form, but one gets logarithmic corrections in an attempt to quantize it, something that is a result of the perturbative 
corrections stemming from the nature of the system, as seen in the corresponding path-integral \cite{IZ}. Moreover, even though the classical action may lack a mass scale, 
such a mass scale is necessarily introduced through regularization. Something similar may apply to a metric: the metric with which we start the analysis of the classical (Hamiltonian) 
system may turn out to be ill-suited when a statistical analysis is performed  due to quantum corrections. This appears to be the case for $\mathcal{S}_q$, as we have indicated in 
 \cite{NK1, NK2, NK3, NK4}. \\

Let us be a bit more specific about this point. The construction and properties of $\mathcal{S}_q$ closely mirror  the corresponding properties of  
$\mathcal{S}_{BGS}$ except when they cannot in order for the entropies not to be identical.   
Usually one starts by considering a particular Hamiltonian describing the 
microscopic system. Its quadratic part induces a Riemannian metric on the configuration / phase space. 
We previously argued in a series of papers \cite{NK1, NK2, NK3, NK4}  that under some additional assumptions, 
this metric may not be best in reflecting the macroscopic properties of the system. We instead argued, without actually  providing concrete 
details, that a more appropriate effective metric would be the ``hyperbolization" of the initially chosen Euclidean metric. 
In this spirit of a modifed/deformed effective metric better reflecting the statistically significant properties of the 
underlying dynamics, the Tsallis entropy (2), and also looking at (9), can be seen as  providing a deformation, or a weighted version, of the micro-canonical density in phase space. 
Then the non-extensive parameter $q$ can be seen as the effective Hausdorff dimension of such a deformation, as was also noticed (alongside its shortcomings) in \cite{NK4}.
The advantage of the present formalism is that it is far more robust than that presented in \cite{NK1, NK2, NK4} as it may very well apply to ``fractal" as well as to 
locally Euclidean spaces. The drawback is that the lack of local regularity of such metrics makes all computations quite a bit harder to perform, as we do not have in our 
disposal local invariants, such as the curvature-derived ones, that may allow us to explicitly compute quantities like (9).                 
                    

\subsection{Some properties of the modulus.}

Proofs of the following properties of the modulus of curve families  and far more general and authoritative discussion can 
be found in   \cite{Vai, Vuor, HeinKosk, Hein, MRSY, HKST, Kosk}.  Here we try to justify/interpret/make plausible, rather 
than derive, their validity from a (Tsallis-) entropic viewpoint: 
\begin{itemize}
   \item {\sf Triviality:} \ To include in our considerations even locally un-rectifiable curves in $\mathfrak{X}$ , 
                      such as snowflakes, we define their modulus to be zero. This is translates into stating that 
                      inaccessible states, due to a conservation law for instance, make no contributions to the entropy.  
  \item {\sf Degeneracy:} \ If the curve family $\mathcal{A}$ contains a constant curve (a point) then there 	are	 
                     no admissible functions $\rho$ and the modulus is infinite. This translates into stating that the entropy 
                     of a certain outcome is equal to 1.    
  \item {\sf Family monotonicity:} For two families of curves obeying $\mathcal{A} \subset \mathcal{B}$ 
              \begin{equation}    
                      \mod_{\!\! q} \mathcal{A} \  \ \leq \!\!\! \mod_{\!\! q} \mathcal{B}
              \end{equation}
                     This states  that a subsystem has less entropy (disorder) than the whole system. 
 \item {\sf Sub-additivity:} Consider a countable set of families of curves $\mathcal{A}_i, \ i=1, 2, \ldots,$ \ \ of \ $\mathfrak{X}$. \ Then
      \begin{equation}  
          \mod_{\!\! q} \left( \bigcup_{i=1}^\infty \mathcal{A}_i \right) \ \ \leq \ \  \ \sum_{i=1}^\infty \!\!\!\! \mod_{\!\! q} \mathcal{A}_i 
      \end{equation}
       This property is  the sub-additivity of \ $\mathcal{S}_q$, \  for \ $q >1$, \  if \  $\mathcal{A}_i$ \ are assumed independent, 
       in the conventional sense of the multiplication of marginals,  of each other. 
  \item {\sf Curve shortening monotonicity:} Let $\mathcal{A}'$, $\mathcal{A}$ be two families of curves in $\mathfrak{X}$ such that each 
                    curve $\gamma \in\mathcal{A}$ has a subcurve $\gamma' \in \mathcal{A}'$. Then 
       \begin{equation}
                 \mod_{\!\! q} \mathcal{A}     \   \   \leq    \!\!\!   \mod_{\!\! q} \mathcal{A}'
       \end{equation}  
       This apparently counter-intuitive statement, becomes more ``reasonable" if one uses an information theoretic interpretation of 
       entropy, as proposed by C. Shannon, and assumes that $\mathcal{A}$ is a letter sequence emitted by a signal source, 
       with $\mathcal{A}'$ being a letter subsequence.                 
\end{itemize}
The above properties show, in a precise sense, that  \! \! $\mod_{\!\! q}$ \  \  is an outer measure on the set of curves of $\mathfrak{X}$. \\


\subsection{Two explicit calculations.}

As may be suspected, it is  quite difficult to explicitly calculate the modulus for a given family of curves. Actual exact calculations are only known 
in very few cases. Usually one seeks upper bounds for $\!\! \mod_{\!\! q}$ \ \  and the far harder to obtain, in general, lower bounds. 
The following two cases are very well-known, but are still instructive and can be considered as possible simplistic examples of a kind of 
``ab initio calculation" of \ \ $\mathcal{S}_q$ \ \ from dynamics. \\ 

\noindent {\large\sf Case 1:} \ \ Consider a cylinder in $\mathbb{R}^n$ endowed with its ($n$-dimensional) Lebesgue measure $\mu_n$ and a family of curves 
$\mathcal{A}$ that joins the two parallel bases $D$ and $D'$ of this cylinder $\mathcal{C}$ which has length $l$. To be more specific, let  
$D\subset \mathbb{R}^n$ be one basis of the cylinder, with $D$ being a Borel set (not necessarily a disk), 
and let the cylinder over it be  
    \begin{equation} 
            \mathcal{C} \ = \     \{ (x_1, \ldots, x_{n-1}) \in D, \ \ \  0 \leq x_n \leq l  \}          
    \end{equation}
Moreover, fix a basis $\{e_i, \ 1\leq i \leq n \}$ of $\mathbb{R}^n$. Then one base of $\mathcal{C}$ is $D$ and the other is $D' = D +le_n$  
Consider the family \ $\mathcal{A}(\gamma)$ \ of all curves \ $\gamma: [a,b]\rightarrow\mathbb{R}^n$ \ with \ $\gamma(t) \in\mathcal{C}, \ \gamma(a)\in D, 
\ \gamma(b)\in D'$.  Then, it turns out that 
\begin{equation} 
    \mod_{\!\! q}\mathcal{A} \ = \  \frac{\mu_{n-1} (D)}{l^{p-1}} \ = \ \frac{\mu_n (\mathcal{C})}{l^p} 
\end{equation}  
We quote the line of reasoning and calculations leading to (14), for completeness. 
The strategy is to find a lower and an upper bound for $\!\! \mod_{\!\! q} \mathcal{A}$ \ \  which coincide, thus establishing the desired equation (14). 
The important observation is that, due to (7), the density $\rho$ can be thought as a kind of inverse ``typical' distance \  $\rho \sim 1/r$. Based on this observation, 
getting an upper bound of $\! \! \mod_{\!\! q}$ \ \ is straightforward: Consider, more generally, a (Borel measurable) set $U\in\mathfrak{X}$ and  a curve family 
\ $\mathcal{A} (\gamma )$ \ in \ $U$. \ Assume that \ $L(\gamma) \geq R > 0$  \ for all \  $\gamma \in \mathcal{A}$. Set 
\begin{equation}
    \rho (x) \ = \ \left\{                               
         \begin{array}{ll}
               1/r,                &         x\in U\\
               0,                  &    x\in \mathfrak{X} \backslash U 
        \end{array}
    \right.
\end{equation} 
Such a  \ $\rho$ \  is an admissible function, i.e. it satisfies (7), hence 
\begin{equation}
   \mod_{\!\! q} \mathcal{A} \ \leq \ \frac{\mu(U)}{R^q}
\end{equation}
As for the opposite inequality of (16), we start by noticing that \ $L(\gamma) \geq l$ \ for all  \  $\gamma\in\mathcal{C}$ \  as \ $l$ \  is the distance between  \ $D$ \ and \ $D'$. \  
Let $\rho$ be an admissible function for all elements $\gamma$ of $\mathcal{A}$. For each point $w$ of the base $D$, consider the (straight) line segment pointing directly 
toward $D'$, which is \ $\gamma_t: [0,l] \rightarrow\mathcal{C}, \ \  \gamma_t = w+te_n$. Let $\mathcal{W} \subset \mathbf{R}^n$ and  $f:\mathcal{W} \rightarrow \mathbb{R}$.  
Recall that for $q \in [1, \infty]$
\begin{equation}
         \| f \|_q \  = \ \left( \int_\mathcal{W} |f|^q \ dvol \right) ^\frac{1}{q} 
\end{equation}
where for $q=\infty$ the integral of (17) is replaced by the essential supremum. Consider $q^\ast \in [1, \infty]$ to be the harmonic (H\"{o}lder) conjugate of $q$, namely 
\begin{equation}
     \frac{1}{q} + \frac{1}{q^\ast} \ = \ 1 
\end{equation} 
The completion of all such $f$ having finite norm (17) forms the Banach space $L^p (\mathcal{W})$.  
Recall \cite{Rudin} that H\"{o}lder's  inequality, which is a functional expression of $q$-convexity,  for such functions \ $f_1, f_2 \in \ L^q (\mathcal{W})$ \ states that      
\begin{equation}
    \| f_1 f_2 \|_1 \ \leq \ \| f_1 \|_q   \    \|  f_2 \|_{q^\ast}
\end{equation} 
Applying this  to (8) we get 
\begin{equation}
   l^{q-1} \int _0 ^l  [\rho(w+te_n)]^q \ dt \ \  \geq  \  \     \left( \int_{\gamma_t} \rho \ dt \right)^ q \ \geq \ 1
\end{equation}
Inserting (20) into Fubini's theorem which is used to integrate over $w\in D$, we find
\begin{equation}
   \mu_{n-1} (D) \ \leq \ l^{q-1} \int_D d\mu_{n-1} \int_0 ^l  [\rho(w+te_n)]^q \ dt \ = \ l^{q-1} \int_\mathcal{C} \rho^q \ d\mu_n  
\end{equation}
which gives that 
\begin{equation}
    \mod_{\!\! q} \mathcal{A} \ \geq \ \frac{\mu_{n-1} (D)}{l^{q-1}}
\end{equation}
Comparing (16) and (22) establishes the validity of (14). For $q=1$ the derivation is similar.  \\

\noindent{\large\sf Comments on Case 1:} \ 
We observe that the modulus of a family of curves for $\mathcal{C}$ is large if the family includes many short $l \ll 1$ curves that are spread out over $D$, and conversely, 
that the modulus is small if the family has few long $l \gg 1$ and ``tightly concentrated" curves over $D$, such as sets of pencils etc. It may also worth pointing out that
$\mod_{\!\! q}$ \ is independent of details of the curves belonging to the family $\mathcal{A}$ such as intersections of the line segments joining $D$ and $D'$ etc.  This insensitivity to 
the details of $\mathcal{A}$ is very advantageous since, it is unlikely that the phase space portrait of a dynamical system described by $\mathcal{S}_q$ will be simple,  even in small 
neighborhoods of the phase space. So it may be approximated by a simpler, yet effectively equivalent phase portrait / family of curves, for which the final results may be both 
equivalent and easier to compute. Such insensitivity to details is at the core of  the basic fact that any entropic functional, be it $\mathcal{S}_{BGS}, \ \mathcal{S}_q$ or something else, 
has to omit (``forget")  many of the details of the underlying system and  take into account only the few features that give rise to an effective macroscopic, 
``thermodynamic", description of the system.\\

\noindent {\large\sf Case 2:} \ \  Let \ $\mathbb{B}_x(r)$ \  indicate the open ball in \ $\mathbb{R}^n, \ n\geq 2$, \ having center $x$ and radius $r$ and \ $\bar{\mathbb{B}}_x(r)$  \ its 
closure. We want to determine the $q$-modulus of the family of all curves \ $\mathcal{G}(\gamma)$ \ having one end on \ $\mathbb{B}_x(r)$ \  and the other end on \
$\mathbb{R}^n \backslash \mathbb{B}_x(R)$ \ for \ $r < R$. \ This is the annular region of radii $r$ and $R$, which can also be physically identified as the interior of the 
spherical capacitor with spherical plates of radii $r$ and $R$, an electrostatic interpretation that we will pursue in Subsection 3.5. 
Let $\mathbb{S}^{n-1} = \partial \mathbb{B}_x(1)$ indicate 
the sphere of unit radius in $\mathbb{R}^n$, and let $\hat{u}$ indicate a radial vector of unit length $\| \hat{u} \| = 1$.   
Consider the radial segment $\gamma_{\hat{u}}: [r, R] \rightarrow \mathbb{R}^n$ defined by $\gamma_{\hat{u}} (t) = t\hat{u}$ and let its arc-length parametrization be indicated by 
$s$. If $\rho$ is admissible for $\mathcal{G}$, then 
\begin{equation}
        1 \  \leq \ \int_{\gamma_{\hat{u}}} \rho \ ds \ \ = \ \int_r^R \rho (t\hat{u}) \ t^\frac{n-1}{q} t^{-\frac{n-1}{q}} \ dt
\end{equation}
H\"{o}lder's inequality (19) applied to (8) gives
\begin{equation}
     \left( \int_r^R [\rho (t\hat{u})]^q \ t^{n-1} \ dt \right)^\frac{1}{q} \ \ \left(\int_r^R t^{-\frac{n-1}{q-1}} \ dt \right)^\frac{q-1}{q} \ \geq  \ 1
\end{equation}
The second factor, call it $C(r,R)$, of the left-hand side of (24) can be readily evaluated giving 
\begin{equation}
 C(r,R) \ = \  \left( \int_r^R t^{-\frac{n-1}{q-1}} \ dt \right)^\frac{q-1}{q} \ \  = \  \   \left\{     
                  \begin{array}{ll}
                                   \left(  \frac{q-1}{|n-q|} \cdot \big{|} R^\frac{q-n}{q-1} - r^\frac{q-n}{q-1}  \big{|}  \right) ^{q-1},  & \mathrm{if} \  \  1 < q \neq n \\
                                                          &   \\
                                  \left( \log \frac{R}{r} \right)^{q-1},     & \mathrm{if} \ \ q=n
                  \end{array}
                             \right.      
\end{equation}
Next, since $\rho >0$ when we integrate over the annulus \ $\mathbb{B}_x(R) \backslash \mathbb{B}_x(r)$  \ and get
\begin{equation}
   \int_{\mathbb{R}^n} \rho^q \ dvol \ \ = \ \  \int_{\mathbb{B}_x(R) \backslash \mathbb{B}_x(r)} \rho^q \ dvol \  \ \geq \ \ \frac{vol (\mathbb{S}^{n-1})}{C(r,R)}   
   \end{equation}
where \ $vol$ \ stands for the volume of $\mathbb{R}^n$ or $\mathbb{S}^{n-1}$ depending on the context.
This provides a lower bound for the modulus $\!\! \mod_{\!\! q}(\mathcal{G})$. \ The fact that this is also an upper bound can be seen by judiciously choosing as an 
admissible function 
\begin{equation}
      \rho(w) \ = \ \frac{|w-x|^{-\frac{n-1}{q-1}}}{[C(r, R)]^{-\frac{1}{q-1}}} \ \cdot \ \chi_{\mathbb{B}_x(R) \backslash \mathbb{B}_x(r)} (w)
\end{equation}
where $\chi_U$ indicates the characteristic function of the set $U$. For $q=1$, choose the admissible sequence of functions
\begin{equation}
    \rho_i (w) \ = \ i \ \chi_{\bar{\mathbb{B}}_x (r+\frac{1}{i}) \backslash \mathbb{B}_x(r)} (w)
\end{equation}
for \ $i\in \mathbb{N}$ \ and then take the limit \ $i\rightarrow \infty$. \ To summarize, we find for the modulus of $\mathcal{G}$ that
\begin{equation}
   \mod_{\!\! q} (\mathcal{G}) \ = \ \left\{
                 \begin{array}{ll}
                           \big{|} \frac{n-q}{q-1}  \big{|}^{q-1} \big{|} R^\frac{q-n}{q-1} - r^\frac{q-n}{q-1}  \big{|} ^{1-q} \ vol(\mathbb{S}^{n-1}), &  \mathrm{if} \  \ q \neq 1, n.   \\ 
                                                 &        \\
                           r^{n-1} \ vol(\mathbb{S}^{n-1}),               & \mathrm{if} \ \ q=1.\\
                                                &         \\
                           \left(\log \frac{R}{r} \right)^{1-n} vol(\mathbb{S}^{n-1}), & \mathrm{if} \ \  q=n.                                              
                 \end{array}
               \right.
\end{equation}\\
\noindent{\large\sf Comments on Case 2:} \ 
We observe that  $\!\! \mod_{\!\! q}(\mathcal{G})$ \ has a power-law behavior for $q \neq n$. This indicates that the entropy $\mathcal{S}_q$ applies to 
systems whose phase space has a power-law dependence on the degrees of freedom, and had been suspected for sometime, but it was explicitly stated in generality 
and some of its consequences were explored in \cite{HT}.   
We also see that for $q=n$ there is a logarithmic dependence on the pertinent parameters having, up to a normalization constant, the form of $\mathcal{S}_{BGS}$.
The reason for this distinct behavior for $q=n$ lies in its conformal invariance, something  which will be discussed more extensively in the next Section. 
The  result of  (29)
provides firmer ground for a geometric interpretation/significance of the non-extensive parameter $q$: this parameter can be seen as the effective dimension of the set of curves 
which are the evolution trajectories of the underlying dynamical system whose statistical behavior is encoded in $\mathcal{S}_q$.  This was previously proposed in    
\cite{NK4} based on constructions on smooth metric measure spaces. Essentially the same interpretation is reached by structures discussed in the present work 
that appear to be quite different from the ones covered in \cite{NK4} and require very few regularity assumptions can be seen as a strengthening of the legitimacy of this 
interpretation.\\

The \ $q=n$ \ case seems to be the most physically relevant. By contrast, and on mathematical grounds alone, $q$ can exceed $n$ without 
any upper bound in the definitions and most proofs related to the modulus of a family of curves.  
In the case \ $q=n$ \ the curves of the family of interest are ``maximally spread" over the phase space. This spread is in part something that the modulus expresses 
quantitatively, even though it is not the only quantititative way of  expressing this fact. To be  more specific about this aspect, consider the metric space 
$(\mathfrak{X}, d_{\mathfrak{X}})$ and let \  $\Omega_1, \Omega_2 \subset \mathfrak{X}$ \  be compact and connected.  Let  
\begin{equation} 
    \Delta (\Omega_1, \Omega_2, \mathfrak{X}) \ =  \{ \gamma : [0, 1] \rightarrow \mathfrak{X}, \ \gamma(0) \in \Omega_1, \  \gamma (1) \in \Omega_2 \} 
\end{equation}
The diameter of a set $\Omega$ in $(\mathfrak{X}, d_\mathfrak{X})$ is defined as
\begin{equation}
        \mathrm{diam}(\Omega) \ = \ \sup \{ d_\mathfrak{X} (x, y), \ \ \forall \ x,y \in \Omega \}
\end{equation}
and a distance function \ $\mathrm{dis} :\mathfrak{X} \times \mathfrak{X} \rightarrow \mathbb{R}_+$ \ 
between \ $\Omega_1, \Omega_2 \subset \mathfrak{X}$ \ can be defined as 
\begin{equation}
        \mathrm{dis} (\Omega_1, \Omega_2) \ = \ \inf \{ d(x,y), \ \ \forall \ x\in\Omega_1, \ y\in\Omega_2 \}
\end{equation}
The relative size of the domain pair \ $(\Omega_1, \Omega_2)$ \  is defined by
\begin{equation}  
     \tau (\Omega_1, \Omega_2) \ = \  \frac{ \min \{ \mathrm{diam}(\Omega_1), \mathrm{diam}(\Omega_2) \} }{\mathrm{dis} (\Omega_1, \Omega_2)}
\end{equation}
It turns out that if \ $\Omega_1, \Omega_2$ \  are disjoint, then there are increasing homeomorphisms \ $\varphi_1, \ \varphi_2: [0, \infty) \rightarrow [0, \infty)$
with \  $\varphi_1(0) = \varphi_2(0) = 0$ \ such that
\begin{equation}
     \varphi_1 (\tau (\Omega_1, \Omega_2)) \ \leq \!\! \mod_{\!\! n} (\Delta (\Omega_1, \Omega_2, \mathfrak{X})) \ \leq \ \varphi_2(\tau (\Omega_1, \Omega_2))
\end{equation}
The optimal functions \ $\varphi_1, \ \varphi_2$ \ are usually quite hard to determine, but their exact form is not particularly relevant for our purposes. 
What is relevant in (34) is to see that \ $\tau(\Omega_1, \Omega_2)$ \ and \  $\!\! \mod_{\!\! n}(\Delta (\Omega_1, \Omega_2; \mathfrak{X}))$ \ are both small or 
both large. So we can state that the relative size which has a very intuitive meaning, captures aspects of the modulus, since (9) seems to be  less 
intuitive to  understand geometrically. \\

One can imagine that at maximal and relatively uniform trajectory spread on phase space, 
the underlying dynamical system should not be ``too different" from an ergodic one. 
The ``not too different" needs to be made precise, and the domain of validity and the conditions under which this statement may be valid need to be checked.  
However this seems to be beyond our current level of understanding and probably not feasible with the available techniques so far as we know, 
so it will not be pursued further in this work. 
The pattern that starts appearing nevertheless,  is that a maximal spread of curves results in a behavior encoded statistically by \ $\mathcal{S}_{BGS}$.  \ The case 
$q < n$ expresses, in a way, a degree of localization of such curves in a lower-dimensional space. 
The statistical behavior of such a system should be  captured by a power law entropy such as \  $\mathcal{S}_q$ \ as (29) suggests. 
Such curve localization to lower dimensional spaces can be quantified via temporal and spatial correlations of the resulting distribution \ $\rho$ \ thus 
providing support, on dynamical grounds, and some credibility to the general belief that systems described by \ $\mathcal{S}_q$ \ have ``fat-tailed" correlations. \\     
       
Then there is the issue of the ubiquitous presence and the reasons for the great success of using \ $\mathcal{S}_{BGS}$ \ in the statistical description of dynamical systems 
of physical, and not only, interest. We presented in a previous work \cite{NK5}, some arguments for the reason of such a success within the limited realm of dynamics 
expressed via random discrete groups having finite presentations. The chaotic hypothesis \cite{GC} is a far broader statement pertaining to hyperbolic systems, within the 
same spirit.  The idea is that ``hyperbolic structures", properly defined and interpreted,  are the ``most" in number or the ``most relevant" (``most" in some general measurable 
sense).  If one or more power-law entropic functionals such as \ $\mathcal{S}_q$ \  describe the ``non-hyperbolic" cases, then the question that arises is which one(s) describe  
which systems. We seem to be quite  far from being able to answering such a question.\\
 
Becoming less ambitious, and based on the above analysis, we may be able to partly address a related question which is also motivated by geometric group theory: 
is there any physically relevant entropic functional of intermediate, namely of slower than exponential but faster than power-law  phase space volume growth ? 
If one thinks in purely terms of discrete groups 
as the kinematic arena of dynamical systems, the question is due to J. Milnor  and the positive answer has been provided by R. Grigorchuk \cite{Grig} via the 
homonymous groups and their generalizations. But from a physical viewpoint, such groups do not seem to have  finite presentations so far as we know, so they may provide 
an answer to physical question  if the dynamics imposes an infinite number of intricately related constraints. It is not clear to us how relevant this may be for 
modelling physical processes.\\

The use of the modulus of a family of curves as one quantitative measure of the spread of the phase space trajectories seems to exclude such an intermediate  behavior 
in the simplest of cases, such as Cases 1 and 2 discussed above. Neither (14) nor (29)  allow for an intermediate growth rate of $\!\! \mod_{\!\! q}$. 
The counter-argument to this statement is that Cases 1 and 2 above, are feasible calculations but of actually quite simple curve families in $\mathbb{R}^n$ 
which may not be expected to be typical of the vast array of  possibilities of the portraits in phase space of dynamical systems. And such a counter-argument would certainly be valid.  
Moreover, someone can question the use of the modulus as a valid, let alone a unique, measure of the complexity of the underlying dynamics, 
despite its functional similarities with $\mathcal{S}_q$, the above arguments non-withstanding. \\

To conclude: even though the modulus of a family of curves approach does not seem to provide any convincing answers to some of the deep 
questions that have been raised during the last 150 years about the dynamical origin and universality of the form of the entropy, it
can be seen as providing a general and flexible enough framework to allow such questions to, at least, be asked in a concrete and mathematically precise form.
Moreover, as is discussed in the next Section, the modulus can address some issues pertaining to the behavior of $\mathcal{S}_q$ that may turn out to be of, 
even substantial, physical significance.  \\

                                                                                                \vspace{0mm}


\section{Invariance of the modulus and implications.} 

One cannot help but wonder, about what is the reason for the different functional dependence on \ $r, R$ \ of $\!\! \mod_{\!\! q}(\mathcal{G})$ between the cases 
\ $q \neq n$ \ and \ $q=n$. \ The behavior of the former is power-law, and of the latter is logarithmic, as seen in (29). What is the significance, if any, of such  a difference?
The answer turns out to be: conformal invariance. In this Section we comment on pertinent aspects of this property, and its generalization/``loosening" to 
``quasi-conformal" invariance which is quite useful for our purposes and is far more flexible, for our purposes, than conformal invariance. Quasi-conformal maps can 
map ``nice" subsets of Euclidean space to even ``fractal" sets (the von Koch snowflake, for instance, is a quasi-conformal image of a circle) and the are quite robust under 
the ``coarse" viewpoint, which amounts to only taking into account large-scale aspects of dynamical systems that survive uniformly bounded changes of metrics. Hence they 
can encode features of systems described by both smooth and fractal spaces, the statistical behavior of some of the latter being conjecturally described by \ $\mathcal{S}_q$.\\


\subsection{Conformal invariance of \ $\mathbf{mod}_{\mathbf{n}}$ } 

Recall that a diffeomorphism between two Riemannian manifolds $(\mathbf{M}_1,\mathbf{g}_1)  \rightarrow (\mathbf{M}_2, \mathbf{g}_2)$ 
is called conformal, if the differential \ $df: T_x\mathbf{M}_1 \rightarrow  T_{f(x)} \mathbf{M}_2, \ \ x\in\mathbf{M}_1,$  \ is a homothety, namely if
\begin{equation}      
    \mathbf{g}_2 (df(x) X, df(x)Y) \ = \ \Lambda(x) \  \mathbf{g}_1 (X,Y)
\end{equation} 
for all  \ $x\in\mathbf{M}_1$ \ and for all \ $X, Y \in T_x\mathbf{M}_1$, and where \ $\Lambda: \mathbf{M}_1 \rightarrow \mathbb{R}$ \  is continuous and $\Lambda (x) > 0$ 
for all \ $x\in\mathbf{M}_1$. \ One can immediately 
see that a conformal map may distort distances but it leaves the angles invariant, at the infinitesimal level. In the linear case, namely if \ 
$\mathbf{M}_1 = \mathbf{M}_2 = \mathbb{R}^n$,\  (35) shows that  $df(x)$ is the multiple of an orthogonal transformation.  From this, one can deduce a synthetic way 
to express that a map is conformal by stating that such an $f$ maps spheres to spheres. \\

 An alternative way to characterize the conformal nature of a map is via its effect on volumes, of the 
afore-mentioned spheres, but also more generally. Consider, for simplicity, $\mathbf{M}_1 = \mathbf{M}_2 = \mathbb{R}^n$ as before,  
and a map \  $f: \mathbb{R}^n \rightarrow \mathbb{R}^n$ \  whose differential is the linear map indicated by $df: T\mathbb{R}^n \rightarrow T\mathbb{R}^n$. \ 
A way of expressing how much a linear operator distorts distances is by using its 
operator norm. In the present case, one way of expressing the operator norm of \ $df$ \  at $x\in\mathbb{R}^n$ and for $Y\in\mathbb{R}^n$ is 
\begin{equation}       
     \| df(x) \| \ = \ \sup_{|Y|=1} \  |df(x) \ Y| 
\end{equation}
Let \ $J_f(x)$ \  indicate the Jacobian determinant of the linear map \ $df$. \  Then \ $f$ \  is conformal if and only if 
\begin{equation}  
     \| df (x) \|^n \ = \ |J_f(x)|
\end{equation}
This can be seen as a local ``compatibility" condition between the linear distortion expressed through the operator norm of \ $df$, \ the (Hausdorff) dimension of 
$\mathbb{R}^n$, and the  volume distortion expressed via \ $J_f(x)$. \ Eq (37) encodes, in a different way, the fact that  locally spheres are mapped to spheres by a conformal map.  
To see the equivalence between (35) and (37), one can observe that in general the linear map \  $df$ \  maps the unit ball centered at the origin $\mathbb{B}_0(1)$ 
to an ellipsoid whose semi-axes have lengths \  $\lambda_1 \leq \lambda_2 \leq \ldots \leq \lambda_n$. \  From the definition of the operator norm  (36), one sees that 
\begin{equation} 
       \| df(x) \| \ = \ \lambda_n 
\end{equation}
and directly from meaning of the Jacobian
\begin{equation}
     | J_f(x) | \ = \ \lambda_1\lambda_2\cdots\lambda_n 
\end{equation}
As pointed out right after (35), a conformal map (35) is locally, in a Riemannian manifold,  a multiple of an orthogonal transformation, therefore  \ 
$\lambda_1 = \lambda_2 = \ldots = \lambda_n$, \ which gives the equality (37). It may be worth noticing that since \ $J_f(x)$ \  is continuous and non-vanishing, the sign of
\ $J_f(x)$ \  remains the same, so one can dispose of the absolute value of the Jacobian in (37) and instead, use the Jacobian itself in that expression.  \\ 

It turns out that the modulus of a curve family $\! \! \mod_{\!\! n}$ \  in a metric  space $\mathfrak{X}$ of Hausdorff dimension $n$ (for such $\mathfrak{X}$ for which 
the relevant concepts make sense such as a manifold $\mathbf{M}$, for instance)  is a conformal invariant. To wit,  if \  $\mathcal{G}$ \  is a curve family in an $n$-dimensional 
domain \ $\Omega \subset \mathfrak{X}$, \ and \ $f: \Omega \rightarrow \Omega'$ \ is conformal, where the dimension of  \ $\Omega' \subset \mathfrak{X}$ \ is also \ $n$, \  then
\begin{equation}
         \mod_{\!\! n} (f\mathcal{G}) \ = \!\!\!  \mod_{\!\! n} (\mathcal{G})
\end{equation}
The justification/proof of  (40) is somewhat instructive, so we present its outline, for completeness: Assume that \ $\rho$  \ is an admissible function for the curve family 
\  $f\mathcal{G}$ \ and let \ $\gamma \in \mathcal{G}$. \  Then 
\begin{equation}
   \int_\gamma \rho (f(x)) \  \| f'(x) \| \  dvol_\Omega \  \ \geq \  \  \int_{f \circ \gamma} \rho \ ds \ \  \geq \  \ 1
\end{equation}   
This plausible (but non-trivial and needing justification) change of variables formula holds for all \ $\gamma \in \mathcal{G}$. \ It shows that \ $\rho (f(x)) \| f'(x) \|$ \ 
is an admissible function for \ $\mathcal{G}$. \ Then (8) implies that   
\begin{equation}
     \mod_{\!\! n} ( \mathcal{G}) \ \leq \ \int_\Omega \  [\rho (f(x)) ]^n \  \| f'(x) \|^n \  dvol_\Omega 
\end{equation}
Using (37) and the change of variables formula gives that 
\begin{equation}
      \int_\Omega \  [\rho (f(x)) ]^n \  \| f'(x) \|^n \  dvol_\Omega  \  =  \ \int_{\Omega'} [\rho (y)]^n \ dvol_{\Omega'} 
\end{equation}
which amounts to stating that 
\begin{equation}
    \mod_{\!\! n} (\mathcal{G}) \  \ \leq \!\!  \mod_{\!\! n} (f\mathcal{G}) 
\end{equation}
Since \ $f$ \ obeys (37) one can see by the symmetry of the argument under the substitution/reversal 
\ $\mathcal{G} \leftrightarrow f\mathcal{G}$ \  that  
\begin{equation}
      \mod_{\!\! n} (\mathcal{G}) \  \ \geq \!\!  \mod_{\!\! n} (f\mathcal{G}) 
 \end{equation} 
thus establishing (40).\\

 Conformal transformations are some of the most studied and understood objects in Geometry. According to a classical theorem of 
Liouville, consider two domains \ $\Omega, \Omega' \subset \mathbb{R}^n, \ n\geq 3$. \ Let  \ $f: \Omega \rightarrow \Omega' $, \  $f\in C^3(\Omega )$  \ be conformal. 
Then \ $f$ \ is the restriction of a M\"{o}bius transformation to \ $\Omega$. \ For proofs of this non-trivial, but well-known fact, one can consult  \cite{Petronio, Iwaniec}.
As a reminder, M\"{o}bius transformations are compositions of Euclidean isometries 
and inversions with respect to round spheres and hyperplanes. See also (55), (56) below. 
The substantially different nature of conformal maps in dimension \ $n=2$ \ should also be pointed out: 
indeed any holomorphic or anti-holomorphic map in 2-dimensions is conformal, in stark contrast to the higher dimensional case. As a result, a ball can only be 
mapped into a ball or into a half-space if $n\geq 3$, something clearly not true for $n=2$.  The group of conformal maps  under composition in $n=2$ is infinite dimensional, 
unlike the higher dimensional case.  This is one important reason for the development and successes of conformal field 
theories in two dimensions which however, largely predictably based on Liouville's theorem, have not been so successfully extended in higher dimensions.\\


\subsection{Quasi-conformal maps and \  $\mathbf{mod }_\mathbf{n}$}

An obvious question is whether,  or to what extent, the results of the previous section apply to the \ $q\neq n$ \ case. Namely, one wonders about 
whether there is one, or a set, of properties of maps that leave invariant \ \  $\!\! \mod_{\!\! q}, \ q\neq n$ \ \ that may somehow resemble the 
conformal invariance  of $\!\! \mod_{\!\! n}$. This is a difficult question to address. On the face of it, one can take a more modest path and 
ask  whether there are more general maps than the conformal ones under which the modulus has ``nice" properties, if not necessarily remain invariant.
We would also like such maps not to be required to be too regular, as someone would like to address issues pertaining to ``fractal" or other general spaces
whose existence provided the motivational ground for the formulation of $\mathcal{S}_q$.   \\

\noindent{\large\sf The ``analytic" definition.} \ \ It should be immediate from the arguments leading to the validity of (40), that it is not easy to 
extend this line of reasoning for the case of $\!\! \mod_{\!\! q}, \ q\neq n$. \ It turns out that it is possible to establish a weaker condition which is called 
quasi-conformal invariance, in such cases. One way to generalize conformal maps is by weakening (37) to homeomorphisms \ $f: \Omega \rightarrow \Omega'$, \
where \ $\Omega, \Omega' \subset \mathbb{R}^n$ \ such that there is a constant \ $K\geq 1$ \ so that 
\begin{equation}  
     \| df(x) \|^n \ \leq \ K \ |J_f(x)|
\end{equation}
Such maps are called $K$-quasi-conformal.
The above (``analytic") definition (46) needs the additional assumption that \ $f$ \ as well as its first derivatives, which appear in \ $J_f$, \ should exist and 
they should be locally integrable. Hence \ $f$ \ should be in the Sobolev space \ $W^{1,1}_{loc} (\Omega, \mathbb{R}^n)$. Moreover, as it is customary 
in measure theory, the requirement is for  (46) to hold almost everywhere in \ $\Omega$. \ It may be worth noticing at this stage, that it is not obvious 
whether \ $f$ \  is almost everywhere differentiable or not. In addition, the requirement that  \ $K\geq 1$ \ is a direct result of linear algebra.\\

 The definition (46) can be 
extended to the case of homeomorphisms \ $f$ \ between Ahlfors regular \  $Q\in \mathbb{R}_+$ \ spaces which are endowed with Borel measures 
$(\mathfrak{X}, d_\mathfrak{X}, \mu_\mathfrak{X}), \ (\mathfrak{Y}, d_\mathfrak{Y}, \mu_\mathfrak{Y})$, still for $K\geq 1$, where 
$f$ belongs to a generalization of Sobolev spaces in the metric measure setting, which are called ``Newtonian spaces" \ 
$N^{1, Q} _{loc} (\mathfrak{X}, \mathfrak{Y})$ \ and where the operator norm in (46) is replaced by a Lipschitz constant \ $\mathrm{Lip}$, \ thus giving    
 \begin{equation}
        [\mathrm{Lip} f(x) ] ^Q \  \  \leq \  \   K \ J_f(x)
 \end{equation}
The Jacobian \ $J_f(x)$ \ in (47) is the Radon-Nikodym derivative of the pull-back measure under \ $f$, \  so it is 
\begin{equation}  
          J_f(x) \   = \    \lim_{r\rightarrow 0} \ \sup_{x\in\mathfrak{X}} \ \frac{\mu_\mathfrak{Y} (f(\mathbb{B}_x(r)))}{\mu_\mathfrak{X} (\mathbb{B}_x(r))} 
\end{equation}
The inequality (47) should hold \ $\mu_\mathfrak{X}$ \ almost everywhere for \ $x\in\mathfrak{X}$.\\

\noindent {\large\sf The ``metric" definition.} \ \ One can immediately observe, by comparing  (37) and (46), that conformal maps are 1-quasi-conformal. Much like conformal 
maps, quasi-conformal 
maps form a group, although it is far harder to determine it when compared to the conformal case. The definition (46) directly generalizes (37). Then one can 
obviously ask whether there is a generalization of  (35) for quasi-conformal maps. The answer is affirmative and it takes place within the general framework 
of metric spaces, without the necessary regularity assumptions implicitly present in (35). For the (``metric") definition, consider the metric spaces 
$(\mathfrak{X}, d_\mathfrak{X}), \ (\mathfrak{Y}, d_{\mathfrak{Y}})$ and let $f: \mathfrak{X} \rightarrow \mathfrak{Y}$ be a homeomorphism. Let 
$x\in\mathfrak{X}$, $r>0$ and consider the two distortion functions
\begin{equation} 
      L_f (x, r) \ = \  \sup \left\{ d_\mathfrak{Y} (f(x), f(y)) : y\in \mathbb{B}_x(r) \right\}  
\end{equation}
and
\begin{equation}
      l_f (x, r) \ = \  \inf  \left\{ d_\mathfrak{Y} (f(x), f(y)) :  y\in \mathfrak{X} \backslash \mathbb{B}_x(r)  \right\} 
\end{equation}
Geometrically, \ $f$ \ distorts the ball \ $\mathbb{B}_x(r) \subset \mathfrak{X}$ \ to some set \ $f(\mathbb{B}_x(r)) \subset \mathfrak{Y}$. \ Then,
\ $L_f (x, r)$ \ is the radius of the smallest ball  centered at \ $f(x)$ \ and circumscribed around the set \ $f(\mathbb{B}_x(r))$. \  Moreover, 
\ $l_f (x,r)$ \  is the radius of the largest ball centered at \ $f(x)$ \ and inscribed in \ $f(\mathbb{B}_x(r))$. \  Consider the ratio, (``linear dilatation") 
\begin{equation}
    K_f (x,r) \ = \ \frac{L_f(x,r)}{l_f(x,r)}
\end{equation}
The homeomorphism \ $f$ \ is \ $K$-quasi-conformal if there exists a positive constant \ $1 \leq K < \infty$ \ such that
\begin{equation} 
     \lim_{r\rightarrow 0} \  \sup_{x\in\mathfrak{X}} \  K_f(x,r) \ \leq \ K 
\end{equation}
In other words, a map is $K$-quasi-conformal when it maps balls to ellipsoids of uniformly bounded eccentricity \ $K$. \ So (52) is the generalization 
of (35) to the $K$-quasi-conformal case. We observe in (52) that since \ $r\rightarrow 0$, \ the definition of quasi-conformality expresses a local property of \ $f$.\\

\noindent {\large\sf The ``geometric" definition.} \ \ There is a third definition (``geometric") for a quasi-conformal homeomorphism \  
$f: \mathfrak{X} \rightarrow \mathfrak{Y}$ \ between two Ahlfors regular \ $Q (\in\mathbb{R}_+)$ \ spaces. It states that \ $f$ \ is \ $K$-quasi-conformal if 
\begin{equation}
       \frac{1}{K} \!\!\!\! \mod_{\!\! Q} (\mathcal{G}) \ \  \leq \!\! \mod_{\!\! Q} (f\mathcal{G}) \ \  \leq \ \  \frac{1}{K} \!\!\!\! \mod_{\!\! Q} (\mathcal{G}) 
\end{equation}  
The condition for \ $f$ \  being $K$-quasi-conformal demands that (53) should hold for all curve families \ $\mathcal{G}$ of $\mathfrak{X}$. \\

The equivalence between the three definitions above (the ``analytic" (46), (47), the ``metric" (52), and the ``geometric" (53)) 
of quasi-conformal homeomorphisms for the case of Euclidean spaces \ $\mathfrak{X} = \mathfrak{Y} = \mathbb{R}^n$ \ holds, but it is not obvious 
and actually quite non-trivial to establish. We refer the interested reader to the literature \cite{Vai, Vuor, MRSY, HKST, Kosk} for this equivalence and 
further information. More generally it is unknown, but extremely unlikely in most cases, that these three definitions are equivalent in a general metric-measure 
space setting: usually one tries to find additional assumptions (such as the doubling condition, the L\"{o}wner condition, the presence of a Poincar\'{e} inequality or 
conditions on the underlying topologies of the metric spaces such as local linear connectedness etc) under which these three definitions are equivalent in 
spaces of interest. \\ 

It should be noticed that quasi-conformal maps have some of the desired properties stated in the introduction of this Section. 
\begin{itemize}
    \item Quasi-conformal maps do not  preserve the Hausdorff dimension of the underlying space. Contrast this with the more familiar case of Lipschitz maps 
            which preserve the Hausdorff dimension of the underlying space even though they may distort absolute distances dramatically. Quasi-conformal maps 
            may distort the relative distances dramatically, hence they are a more ``flexible" set of maps than the Lipschitz ones. Changing the Hausdorff dimension 
             may be considered, for a class of maps, to be anything from mildly annoying to probably sufficient 
            to rule them out as  physically interesting or to even declare them outright irrelevant. Our view is the exact opposite: despite the substantial  geometric 
            consequences 
            that such a flexibility may entail, it also allows us to treat in a unified matter very regular ``Euclidean" and very irregular ``fractal" spaces. It is well-known,
            for instance, that the image of the unit circle  $\mathbb{S}^1$ under a quasi-conformal map (called a ``quasi-circle") can be the von Koch snowflake curve, 
            which possesses no rectifiable (``finite length") sub-curves. A general characterization of quasi-circles seems to be well-known. 
            Far less is known for the images of higher-dimensional spheres under quasi-conformal maps. However, the fact is that even in higher dimensions 
            quasi-conformal maps can change the Hausdorff dimension of the underlying space. Our view is that such a dimension change should not be considered an 
            indictment of the relevance or significance of quasi-conformal maps for Physics, but a realization of the limitations of the concept of Hausdorff dimension 
            when one includes in the analysis ``fractal" spaces or takes a coarse view of these structures.  
            Indeed, there have been numerous definitions of  so-called ``dimensions" over the decades, most of which are constructed to agree on Euclidean
            spaces, but to give different results for less regular (more ``exotic")  spaces. Many of such definitions have been motivated by 
            studies  dynamical systems \cite{Pesin} or large-scale (``coarse") geometry \cite{Gromov} and should be more relevant / appropriate for our purposes.   
    \item It follows from the ``geometric" definition (53): quasi-conformal maps
             may change distances in a very ``dramatic" way, but still the modulus of a family of curves does not change all that much if viewed at a large, ``coarse", scale. 
             Even though coarse aspects of phase space may be important for understanding irreversible processes, for the quantization of the underlying system 
             if one needs to do that etc,  their significance is unclear, to us at least, for the underlying classical dynamics. It is entirely possible that geometrically ``small" sets, 
             such as attractors, have considerable significance in the dynamical and statistical description of a system, so ignoring such small-scale details as the 
             quasi-conformal maps do may not be prudent for the analysis of a particular system. On the opposite side of this, some level of coarse-graining inevitably 
             creeps in the definition of any  entropic functional, and in that respect, maybe some  quasi-conformal maps (probably with additional conditions imposed 
             to reduce the resulting set of acceptable maps for our purposes) may turn out to being useful in ``naturally" performing and controlling such a level of coarse-graining.    
    \item Quasi-conformal maps turn out to be almost everywhere differentiable. This is quite non-trivial to establish, but the situation is quite similar to the 
             case of the Lipschitz maps. The implications are obvious: we have available the first order calculus that we are somewhat familiar with, and therefore 
             can differentiate a map almost everywhere etc. So many of the familiar and intuitive in Physics properties of maps are applicable.    
    \item Here are three examples of quasi-conformal maps that may help someone get an intuitive sense about these structures:
              First of all, any conformal map is 1 quasi-conformal. Second, the planar map \ 
              $h: \mathbb{R}^2 \rightarrow \mathbb{R}^2$ \ given by \  $h(x,y) = (x,2y)$ \  is 2 quasi-conformal. 
              Third, a standard example (but not obvious) and potentially more interesting for Physics is that of the radial stretching \  
              \  $f:\mathbb{R}^n \rightarrow \mathbb{R}^n$ 
             \begin{equation}
                 f(x) \ = \ x |x|^{\alpha - 1}, \ \hspace{15mm} \alpha \ > \ 0   
             \end{equation}             
\end{itemize}


\subsection{Quasi-conformality and quasi-M\"{o}bius maps.} 

As noted in Subsection 3.1, Liouville's theorem characterizes conformal maps of $\mathbb{R}^n$ as a subset of M\"{o}bius transformations. The question
that naturally arises is whether there is a similar characterization for quasi-conformal maps. The answer is affirmative in the following way. First of all, we have 
to recall that the M\"{o}bius maps in \ $\mathbb{R}^n \cup \{ \infty \} $ \ are generated by 
\begin{itemize}
   \item Reflections $f_1$ with respect to the hyperplane \ $H_t (u) \ = \ \{ x\in\mathbb{R}^n : x\cdot u = t \} \cup \{ \infty \} $, \  given by
                \begin{equation} 
                       f_1(x) \ =  \ x - 2 \frac{u}{|u|^2} \ (x\cdot u -t),   \hspace{15mm} f(\infty) \ = \ \infty
                \end{equation}
            where \ $\cdot$ \ stands for the usual  inner product in $\mathbb{R}^n$, \ $u \in\mathbb{R}^n \backslash \{ 0 \}$ \ and \ $t\in\mathbb{R}$.    
   \item Inversions $f_2$ with respect to the sphere $\mathbb{S}_{x_0}^{n-1}(R)$ centered at $x_0\in\mathbb{R}^n$  of radius $R>0$,  \ given by           
               \begin{equation}
                  f_2(x) \ = \ x_0 + \frac{(x-x_0) R^2}{|x-x_0|^2}, \hspace{10mm} f_2(x_0) = \infty, \hspace{10mm} f_2(\infty) = x_0
               \end{equation} 
\end{itemize}    
Following (55), (56) one can see that the conjectured symmetries (3), are a generating set of the M\"{o}bius maps in \ $\mathbb{R}^2 \cup \{ \infty \}$. \ 
However, we should notice that (3) are conjectured to be invariances under M\"{o}bius maps  of the non-entropic parameter $q\in\mathbb{C}$, rather than 
be directly applicable to $\mathcal{S}_q$ itself, which is what the above analysis indicates. The transformations of $q$ given by (3) are two generators of the 
Lie group $PSL(2, \mathbb{C})$ which is also the group of orientation-preserving isometries of the 3-dimensional hyperbolic space.  \\

In higher dimensions, the generalization relies of the M\"{o}bius maps relies on the familiar, from the classical Euclidean and hyperbolic geometries, 
concept  of the cross-ratio \cite{Petronio}. Let \ $(\mathfrak{X}, d)$  \ be a metric space and consider four distinct points \ $x,y,z,w \in \mathfrak{X}$.
Then the cross-ratio of \ $x,y,z,w$ \ is defined as 
\begin{equation}      
       [x,y,z,w] \ = \ \frac{d(x,z) \ d(y,w)}{d(x,y) \ d(z,w)}
\end{equation}
The motivating property, for our purposes, is that if \  $f: \mathbb{R}^n \cup \{ \infty \} \rightarrow \mathbb{R}^n \cup \{ \infty \}$ \ is a M\"{o}bius map,
 then it leaves the cross ratio invariant, namely
\begin{equation}
     [f(x), f(y), f(z), f(w)] \ = \ [x,y,z,w]     
\end{equation}
This provides a  characterization of the underlying geometric features of \ $(\mathfrak{X}, d)$ \  in terms of properties of its automorphism group,
in the spirit of F. Klein's Erlangen program. \\

Based on the above, consider two metric spaces \ $(\mathfrak{X}, d_\mathfrak{X}), \ (\mathfrak{Z}, d_\mathfrak{Z})$  \ and \  $\eta :[0, \infty) \rightarrow [0, \infty)$ \ 
a homeomorphism. An injective map \  $f: \mathfrak{X} \rightarrow \mathfrak{Z}$ \  is \ $\eta$-quasi-M\"{o}bius \cite{Vaisala2}, if for every quadruple of distinct points 
\ $x,y,z,w$ \  one has 
\begin{equation} 
   [f(x), f(y), f(z), f(w)]_\mathfrak{Z} \ \leq \  \eta ([x,y,z,w]_\mathfrak{X})   
\end{equation}
The corresponding relevant property is that quasi-M\"{o}bius maps of metric spaces are quasi-conformal. 
It is also notable that quasi-M\"{o}bius maps, even though they can distort distances tremendously, they are continuous. 
Moreover, even though quasi-M\"{o}bius maps do not leave invariant the Hausdorff dimension of the 
underlying space, they do leave invariant other quantities of potential interest, such as the conformal dimension (by definition), and the Assouad-Nagata dimension 
\cite{Xie}, for instance. This may strengthen somewhat our comment above that other dimensions, 
different from the Hausdorff one, may be appropriate for analyzing systems whose collective behavior is described by $\mathcal{S}_q$.\\    

There is, of course, the question of how close to or different from the quasi-M\"{o}bius maps are with respect to the  quasi-conformal ones. 
We know, for instance, that for round spheres \ $\mathbb{S}^n$ \ of any finite dimension \ $n \geq 3$ \ the groups of conformal and M\"{o}bius maps are isomorphic. 
In the ``quasi-" case, the respective two classes of maps are equivalent for ``nice" spaces admitting a first order calculus, such as $\mathbb{R}^n$, Heisenberg groups 
endowed with Carnot-Carath\'{e}odory metrics etc. For more general spaces very little is known. 
For further information, details and generalizations about the quasi-conformal and quasi-M\"{o}bius maps, one may wish to consult \cite{Vaisala3} 
among several other references. \\     


\subsection{Quasi-conformality   and  Muckenhoupt weights.} 
   
 In \cite{NK6} we presented properties of \ $\mathcal{S}_q$ \ conjecturally stemming from the long-range interactions of the underlying dynamical system.  
 We expressed aspects of long-range interactions  through the formalism of Muckenhoupt weights which plays a significant role in Real Variables / Harmonic Analysis
 and Potential Theory \cite{Stein}. In this subsection, we would like to point out that quasi-conformal maps fit well with aspects of the theory of weights that 
 may be pertinent to our goal of providing a dynamical base and range of applicability of \ $\mathcal{S}_q$, \ 
 thus providing a connection with the  treatment of \cite{NK6}. \\  
   
 We refer to \cite{Stein} for an excellent exposition of Muckenhoupt weights. In this work we only need the following. We use the very common notation, 
 for \   $f: U \rightarrow \mathbb{R}$ 
 \begin{equation}
         \fint_{U} f  \ = \ \frac{1}{vol \ U} \int_U  \ f \ dvol_U
 \end{equation}
 where \ $U\subset \mathbb{R}^n$, \ is measurable with \  $0 < vol \ U < \infty$. \ The measure used in (60) is the Lebesgue measure of $U$.  
 Let \ $w$ \ be a locally integrable function on $\mathbb{R}^n$, or more precisely \ 
 $w\in L^1_{loc} (\mathbb{R}^n)$. \ Then \ $w$ \ is called an \ $A_p, \ 1 < p < \infty$ \ (Muckenhoupt) weight,  if for all balls \ $\mathbb{B} \subset U$
 \begin{equation}  
        \left( \fint_\mathbb{B} w^p \right)^\frac{1}{p} \leq \ C(p,w) \left( \fint_\mathbb{B} w^\frac{1}{1-p} \right)^{1-p}
 \end{equation}
 and for \ $p=1$
 \begin{equation}
       \fint_\mathbb{B} w \ \leq \ C(1,w) \ \mathrm{essinf}_\mathbb{B}w 
 \end{equation}
 where \ $\mathrm{essinf}_{V}$ \ indicates the essential infimum of its argument in the set \ $V$. \  Set 
 \begin{equation}
       A_\infty \ = \ \bigcup_{p>1} A_p
 \end{equation}
 and notice the inclusion \ $A_1 \subset A_q \subset A_p$ \  for \  $1 \leq q \leq p$. \ One can then prove that \ $w\in A_\infty$ \ is equivalent to the 
 existence of \ $q>1$ \ and a constant \ $C$ \  such that 
 \begin{equation}
     \left( \fint_\mathbb{B} w^q \right)^\frac{!}{q} \ \leq \ C \fint_\mathbb{B} w
 \end{equation}  
 for all balls \ $\mathbb{B} \subset U$. \  To fix the notation again, let \ $U, V \subset \mathbb{R}^n$, \  $x\in U$ \ and let \ $f: U\rightarrow V$ \ be a 
 homeomorphism. Set
 \begin{equation} 
         \nu_f (x) \ = \ \lim_{r\rightarrow 0} \frac{vol \ f(\bar{\mathbb{B}}_x(r))}{vol \ \mathbb{B}_x(r)} 
\end{equation}     
In the case that \ $f: U \rightarrow V$ \ is quasi-conformal and \ $n\geq 2$, \ then \ $vol  \ f(\Omega) = 0$ \ if and only if \ $vol \ \Omega = 0$ \ for 
\ $\Omega \subset U$ \ measurable. This is  property, interpreted in terms of the deformations of an elastic medium  means 
that a hole in such a medium is mapped into a hole via a quasi-conformal map. This clearly does not need to be true in phase space, as there 
is no physical principle requiring or ruling out such a behavior of maps. Moreover
\begin{equation}
     vol \ f(\Omega) \ = \ \int_\Omega \ \nu_f(x) \ dvol_\Omega
\end{equation}   
  with \ $\nu_f(x) > 0$ \ almost everywhere and \ $\nu_f \in A_\infty$. It should also be mentioned that comparing (48) and (65), one can prove that
  \begin{equation}
     |J_f(x)| \ \leq \nu_f(x)    
  \end{equation}
  for \ $f\in W^{1,2}_{loc} (\Omega, \mathbb{R}^n)$. \  The above relation of quasi-conformal maps to $A_\infty$ may be considered too arcane for a
  physical interpretation. However it becomes much more relevant, for our purposes, if we recall the following property of such \ $A_\infty$ \ weights:
  let $\mathbf{Q}$ denote a cube of side length of 1 unit. Assume that \ $U \subset \mathbf{Q} \subset \Omega \subset \mathbb{R}^n$. \ 
  Then there are constants \ $C$ \ and \ $\alpha$ \ such that \ $f: \Omega \rightarrow \mathbb{R}^n$ such that
  \begin{equation}     
     \frac{\int_U w}{\int_\mathbf{Q} w} \ \leq C \left( \frac{vol \ U}{vol \ \mathbf{Q}} \right)^\alpha
  \end{equation}
  Since the volume distortion function $\nu_f(x)$ is an $A_\infty$ weight, it satisfies (68). This is different way of stating that  
  quasi-conformal maps do not distort the relative volumes too much.\\
  
  The most important, for our purposes, conclusion from (68) is that quasi-conformal maps distort the volumes in a power-law way. 
  Let's recall a conclusion of  \cite{HT}  that $\mathcal{S}_q$ describes systems whose phase space volume increases in a power-law manner. 
  We are certainly  interested in maps that preserve this power-law dependence, if we want to understand the dynamical origins of $\mathcal{S}_q$. 
  It is quite important to know that even though quasi-conformal maps may distort lengths very substantially (after all the von Koch snowflake curve 
  is a quasi-circle as has been pointed out above), they have a far more controlled effect
  on relative volume distortions and such a relative distortion remains within the general class of power-law functional dependence.  
  Hence, aspects of the predictions given by \ $\mathcal{S}_q$ \ may change under quasi-conformal maps, but many qualitative, or even 
  quantitative, features of such systems will change but relatively uniformly, i.e. not in a dramatic way. \\
  
  We can also conclude that, alongside reconsidering the role of the Hausdorff dimension of the underlying space, one may wish to also widen the
  class of maps of interest from conformal ones to quasi-conformal, quasi-M\"{o}bius or even more generally to quasi-regular maps, when discussing the underlying 
  dynamics of systems whole collective behavior is encoded by \ $\mathcal{S}_q$. \  This allows us to treat systems having a ``regular" / locally Euclidean 
  phase space and some ``fractal" phase spaces in a unified manner, and \  $\mathcal{S}_q$ \ can still be applicable in describing both classes of such systems. \\                 
     

\subsection{Quasi-conformality and capacities.} 

 Even though the concept of the modulus of a curve family appears somewhat arcane to a physicist, its roots can be traced back to developments of the 18th, 
 and primarily the 19th, century in the theory of electricity and magnetism. This has already been mentioned above a couple of times, we intend to elaborate  
 on this connection in this Subsection. \\
 
 Consider a simple capacitor, as in electrostatics, each of the electrodes of which \ $\Omega_1, \  
 \Omega_2$\  is held at potentials \ $\Phi_1, \ \Phi_2$ \ respectively. To make things a bit more realistic, one may wish to take into account the finite conductivity of 
 these electrodes and also consider variable in time electromagnetic fields. To simplify the latter, we wish to consider only slowly varying ``quasi-stationary"
 electric fields as this allows us to ignore the effects of the existence and variations of the magnetic fields and of the emitted electromagnetic radiation and 
 pretend that the situation can be described well enough by static, hence conservative, fields. This way, we do not have to solve the full set of Maxwell's equations to
 determine the fields, but we can instead use purely electrostatic concepts, thus ignoring the vector potential etc. To take into account the finite conductivity and the 
 inevitable dielectric properties of the electrodes, let's assume that the scalar potential of the electrodes  satisfies
 \begin{equation}                                   
      \Phi (\Omega_1) \leq 0, \hspace{15mm} \Phi (\Omega_2) \geq 0                         
 \end{equation}                            
We do not have to confine ourselves to a treatment where the electrodes \ $\Omega_1, \ \Omega_2$ \ are in \ $\mathbb{R}^n$, \  but can be more general and assume 
that they are in some manifold, or even a more general metric measure space \ ($\mathfrak{X}, d_\mathfrak{X}, \mu$) \ endowed a distance function $d_\mathfrak{X}$ 
and a Borel regular measure \ $\mu$. \ Then, in a Euclidean setting, the total electrostatic energy contained in this capacitor is defined and it is given by 
  \begin{equation}                               
                   \mathcal{E} (\Omega_1, \Omega_2, \mathbb{R}^n) \ =  \  \int_{\mathbb{R}^n} \vec{E}\cdot \vec{E} \ dvol                 
  \end{equation}                           
  as is well known from Maxwell's theory, where \ $\vec{E}$ \ stands for the electric field and the integration is over the whole space of \ $\mathbb{R}^n$. \ 
  In case we ignore fringing effects, and the set of the two 
  electrodes have a ``nice" enough shape, we usually tend to ignore the electric field outside this capacitor and therefore confine the domain of integration to the 
  ``interior"  of such a capacitor. It is notable that the electrostatic energy expression (70) does not actually involve the electric field \ $\vec{E}$ \ itself but only its 
  magnitude $|\vec{E}|$. This allows the generalization of (70) to general enough metric spaces \ $\mathfrak{X}$ \ where the concept of an ``angle" may not be defined. 
  To be slightly more general, one can consider the 
  energy functional
  \begin{equation}     
       \mathrm{cap} (\Omega_1, \Omega_2, \mathbb{R}^n) \ = \ \int_{\mathbb{R}^n} |E|^q \ dvol
  \end{equation}
  We have renamed this functional, since``capacity" expresses better what we compute, namely the capacitance of the capacitor under consideration. 
  Let \ $1\leq q <\infty$ \  be a free parameter. In the familiar 
  context of electrostatics \ $\vec{E} = - \nabla\Phi$. \ However, for the more general metric measure space \ $\mathfrak{X}$ \ the concept of direction is generally 
  not well-defined. But observing (71), one sees that to define a capacity for a ``capacitor" in \ $\mathfrak{X}$ \ it is sufficient to just consider/define the magnitude 
  of the gradient of the potential, rather than the gradient of the potential  itself, something which would be ill-defined. For a conservative field, in \ $\mathbb{R}^n$ \  
  we have that  
  \begin{equation}
       \Phi (x_2) - \Phi (x_1) \ = \  \int_\gamma \nabla\Phi  \  ds
  \end{equation}
for any curve \  $\gamma$ \  with end-points \ $x_1 \in \Omega_1, \  x_2\in\Omega_2$, \  arc-length parametrized by \ $s$, \  with \ $\gamma \in \Delta (\Omega_1, 
\Omega_2, \mathbb{R}^n)$ \ in the notation of (30). To generalize (72) to more general \ $\mathfrak{X}$, \  one immediately sees that  
\begin{equation}
    |\Phi (x_2) - \Phi (x_1)| \ \leq \ \int_\gamma |\nabla\Phi| \ ds   
\end{equation}
and proceeds as follows \cite{HeinKosk}: In the terminology of \cite{HeinKosk}, consider a function \ $f: \mathfrak{X} \rightarrow \mathbb{R}$ \ and a Borel function 
\ $F: \mathfrak{X} \rightarrow \mathbb{R}_+$. \ Then \ $F$ \  is called an upper gradient of \ $f$ \  if
\begin{equation}
     |f(x_2)  - f(x_1)| \ \leq \  \int_\gamma F(\gamma(s)) \  ds
\end{equation}
for any rectifiable curve \ $\gamma$ \ in \ $\mathfrak{X}$, \ arc-length parametrized, with endpoints \ $x_1, \ x_2$. \ It follows that there is no reason why upper 
gradients should be unique, and actually they never are.  Obviously \ $F$ \  being infinity is an upper gradient of any function \  $f$. \  If \ $\mathfrak{X}$  \ 
has no rectifiable curves, (example: if \ $X$ \ is  the von Koch snowflake curve), then \ $F=0$ \ for any function in \ $\mathfrak{X}$. \ 
There are several slightly different definitions, terminology  and constructions related to this concept in the literature; in this work we follow \cite{HeinKosk}. 
Alternative definitions with their consequences and more recent developments can be found in \cite{HKST}.  Going back to the capacities (71), consider now 
\ $\Omega_1, \ \Omega_2 \ \subset \ \mathfrak{X}$ \ and again let \ $1\leq q < \infty$. \ Then the $q$-capacity of the pair of ``electrodes" \ $\Omega_1, \ \Omega_2$ \ is 
\begin{equation}
        \mathrm{cap}_q (\Omega_1, \Omega_2, \mathfrak{X}) \ = \ \inf_\Phi  \int_\mathfrak{X} \Phi^q \ d\mu
\end{equation}
where the infimum is taken over all upper gradients \ $\Phi$ \ of all functions \ $f: \mathfrak{X} \rightarrow \mathbb{R}$ \ such that \ $f|_{\Omega_1} \leq 0$ \ and \ 
$f|_{\Omega_2} \geq 1$.\  One can easily see that the set of upper gradients of a function \ $f$ \ is convex.\\

The preceding definitions and discussion about capacities is of relevance to the present work because there is a fundamental equality, stating that \cite{Hein}
\begin{equation}   
   \mathrm{cap}_q (\Omega_1, \Omega_2, \mathfrak{X}) \ = \ \bmod_q (\Omega_1, \Omega_2, \mathfrak{X})
\end{equation}
which equates the somewhat ``exotic" definition of the modulus of a curve family to the far more familiar, to a Physicist,  definition of capacity of a capacitor.\\

A potential objection to the above analysis is that in classical electromagnetic theory all fields are considered to be sufficiently smooth, almost everywhere.
A direct implication of Maxwell's equations (or Gauss' law, or Poisson's equation in electrostatics) is that the electric field lines, which are the integral curves 
of the electric field, cannot intersect in any way, either transversally or not. The definitions of the $\bmod_q$  and $\mathrm{cap}_q$ do not place such a 
restriction in selecting the members of a curve family $\mathcal{G}$. On geometric grounds, the best understood modulus is the one having an index equal to
the ``effective dimension",  of the configuration/phase space of the underlying dynamical system. For this case, assume that the underlying space is either 
Ahlfors $Q$-regular or, even more generally, its measure  obeys
\begin{equation} 
     \mu (\mathbb{B}_x(r)) \ \leq \ Cr^Q
\end{equation}
it turns out that the set of curves passing through a particular point \ $x\in\mathfrak{X}$ \ has a $Q$-modulus which is equal to zero \cite{Hein, HeinKosk}. 
Therefore whether the curves  (``electric field lines") of the family \ $\mathcal{G}$ \ intersect each other or not, does not really make any difference for 
the calculation of  \ $\bmod_Q (\mathcal{G})$ \ and hence such an issue can be safely ignored. However it should be noticed that this conclusion is no longer 
true in the case of \ $\bmod_q$ \ for \  $q>Q$. \\

There is also a more general issue of the regularity of the curves belonging to the family \ $\mathcal{G}$ \ that we 
are considering. From the viewpoint of classical electromagnetic theory, all field lines are almost everywhere differentiable: one has to allow for lack of 
differentiability  or even occasional discontinuities of the field lines to accommodate lower dimensional (sheets, filaments, point-like charges etc) 
distributions of charges and currents. The question that arises is whether such regularity requirements of the (electric) field lines invalidate the generality 
of the definitions and of the arguments leading to (76). The answer to such question is, in general, affirmative. However, if one adds some ``reasonable", 
from a physical viewpoint and certainly shared by the usually considered subsets Euclidean spaces,  
 properties of \ ($\mathfrak{X}, d_\mathfrak{X}, \mu$) \  such as being compact and locally quasi-convex
 then it turns out that one can restrict the class of functions used in (75) to Lipschitz ones. This is convenient as Lipschitz functions are almost everywhere 
 differentiable, effectively bringing us back to the familiar setting of first-order calculus employed in electricity and magnetism.\\        

The above arguments are somewhat suggestive of the fact that \ $\bmod_q$ \ and its properties have something to do with the calculation \ $\mathcal{S}_q$ \ 
when the underlying dynamical system has evolution trajectories which are the members of the family of curves \ $\mathcal{G}$ \  under investigation. Therefore
the calculation of such capacities with methods familiar from electrostatics may provide some insight into the dynamical foundations of \ $\mathcal{S}_q$ \ 
in concrete cases. Such explicit analytic calculations are sorely lacking. We such a state of affairs as a major impediment in the further development of the 
foundations of \ $\mathcal{S}_q$ \ and of the several proposed non-additive entropies, in general.  We  hope that relations such as (76) may help address
such open issues.



\section{Conclusions and outlook.}

 
In the present work we have presented comments on the apparent similarity between the Tsallis entropy \ $\mathcal{S}_q$ \ seen from a micro-canonical 
viewpoint and the modulus of a curve family of the underlying dynamical system. This is part of an approach that attempts to determine the dynamical foundations 
$\mathcal{S}_q$, or other power-law entropies, and at the same time elucidate the reasons for the wide applicability and success in describing such a wide range of phenomena, of 
the Boltzmann/Gibbs/Shannon entropy $\mathcal{S}_{BGS}$.  \\

We have seen that even though the modulus of a curve family has a different origin and flavor from that of entropy, it can at least formally reproduce some of the desired
properties of \  $\mathcal{S}_q$. \ We presented arguments about the origins of the possible and suspected underlying conformal invariance expressed through the entropic 
indices  of some systems. We commented on the role of quasi-conformal transformations as a set of maps whose properties are maintained  for very regular ``Euclidean" 
spaces  and for some irregular ``fractal" spaces which had helped motivate the introduction of \ $\mathcal{S}_q$. \ We also presented an electromagnetic/field-theoretical
interpretation of the modulus of a family of curves as the capacitance of a capacitor whose two plates are the endpoints of the members of this family of curves.\\

There are several possible next steps in this direction of work: one of them is to check to what extent the power-law entropies can be seen as objects living on the 
boundaries of configuration/phase spaces that are Gromov hyperbolic, whose underlying dynamics is described by $\mathcal{S}_{BGS}$. 
This is clearly motivated by the ideas and the  developments related to the Mostow rigidity of lattices and symmetric spaces. 
 Another direction is to attempt to elucidate the origin of the entropic 
parameter \ $q$ \ whose origins are still not particularly well understood and whose ab initio calculation in systems with many degrees of freedom appears to be currently 
intractable. One can view \ $q$ \ as quantifying a measure-theoretical analogue of metric ``snow-flaking". In this spirit, and being motivated by the underlying quasi-conformal
properties  and the concomitant  ``coarse" description of the 
underlying dynamical system, one can naturally inquire about the significance for $\mathcal{S}_q$ of the conformal, rather than the Hausdorff, dimension of the configuration 
or phase space. We will pursue both of these, and possibly other approaches in this spirit, in the immediate future. 



                                                                     \section{Acknowledgements.}

                                                                                       \vspace{0mm}

We would like to thank Professor Constantino Tsallis for providing some historical background and references regarding the origins of the 
Tsallis entropy $\mathcal{S}_q$. We are grateful to the referee for his/her feedback that has helped  improve considerably the clarity of the exposition.




\end{document}